# A Risk-driven Model for Work Allocation in Global Software Development Projects

Ansgar Lamersdorf
University of Kaiserslautern
Software Engineering Research
Group, PO Box 3049
67653 Kaiserslautern, Germany
ansgar@lamersdorf.net

Jürgen Münch
University of Helsinki
Department of Computer Science
P.O. Box 68
(Gustaf Hällströmin katu 2b)
FI-00014 Helsinki, Finland
juergen.muench@cs.helsinki.fi

Alicia Fernández-del Viso Torre,
Carlos Rebate Sánchez
Indra Software Labs
Madrid, Spain
{afernandezde; crebate}
@indra.es

*Abstract*—**Risks and potential benefits of distributing software development projects globally depend to a large extent on how to allocate work to different development sites and regions. Existing methods in task allocation are likely to omit the relevance of considering a multitude of criteria and the impact of task distribution on risks and potential benefits. To assess risks stemming from specific work distributions and to exploit organization-specific experience, we have developed a customizable risk-driven model. It consists of two main steps: Suggesting a set of task allocation alternatives based on project- and site-specific characteristics and analyzing it with respect to possible project risks stemming from the work distribution. To evaluate the model, we conducted a series of semi-structured interviews in a multinational IT company. The results of the evaluation show that the suggestions of the model mostly comply with the retrospective views voiced by the involved experienced managers.**

*Keywords: Global Software Development, Task Allocation, Risk Management, Project Planning*

## I. INTRODUCTION

The distribution of software development work over several sites, which may be spread across different countries, has become a common practice in industrial software engineering [1]. It is mostly driven by various expected benefits such as labor cost reductions, access to resources, or proximity to markets. These expected benefits (especially the labor cost savings) mainly address strategic organizational goals. However, on a tactical project level, global software development (GSD) imposes a set of risks and problems such as reduced productivity, lack of trust, or rework needs [2-4], which can threaten project success and, in consequence, the strategic goals. These problems can generally be traced back to two main causes:

- Insufficient abilities (e.g., absence of domain knowledge) [3] or problematic characteristics (e.g., high turnover rates, loss of intellectual property) at certain sites (often especially at low-cost sites [5]), and

- problems at the interfaces between two distributed sites, caused by language, cultural, time zone, and other barriers [6-8].

Both causes are impacted by the decision on work allocation (i.e., the distribution of work packages to sites): Assigning, for example, two closely coupled tasks to sites with high barriers can largely increase communication problems between them. On the other hand, the potential benefits of GSD are also impacted by the work allocation if, for example, more work is assigned from high-cost to low-cost sites, resulting in a reduction of overall costs.

In practice, it can be seen that, after a phase of focusing on labor cost rates alone [5], work is often allocated using simple strategies such as nearshoring [9] or the use of bridgehead sites [10]. However, work allocation in industry is still mostly done unsystematically and only focuses on few criteria such as cost rates, availability, and expertise [11]. In our view, the typical risks of GSD projects and their causes are not enough considered, even though experienced practitioners in GSD are aware of them and can tell failure stories [12]. Moreover, work allocation is typically done ad-hoc and not in a transparent manner, without any defined process or guidelines. This makes the quality of the allocation decision (and thereby its impact on project goals) solely dependent on the individual expertise of the decision makers involved. These problems are also not sufficiently covered by existing task allocation methods.

We address these problems via two contributions:

- We propose a tool-supported integrated assignment model and an accompanying process for the development of custom-tailored models and their usage. It includes a stochastic sub-model [13] and a risk identification sub-model [14]. The contribution consists of the integration of both models into a coherent model for supporting systematic work allocation decisions and the definition of the accompanying processes.

- We evaluated the integrated assignment model in a company context. The evaluation considers the

assignment suggestions of the integrated model. The risk identification capabilities were evaluated in a previously performed study [14].

The article is structured as follows. Section 2 gives an overview of related work in experienced-based risk identification in GSD and decision support in GSD work allocation. Section 3 explains the overall model in detail, followed by a presentation of the processes for model development and usage in Section 4. Section 5 presents the evaluation of the stochastic assignment model at Indra Software Labs. Finally, Section 6 concludes the article and discusses limitations as well as future work.

## II. RELATED WORK

Many studies analyze current practices of global and distributed software development. Most describe specific risks and problems of globally distributed work and name strategies and tactics applied in practice to overcome the problems. Such problems are, for example, described by Herbsleb and Mockus [2], who analyze the impact of distribution on productivity and the underlying causes, by Herbsleb et al. [3], who report on the experiences made in GSD projects at Siemens, by Smite and Moe [4], who describe the impact of lack of trust between GSD teams, and in several reports on specific GSD projects [6-8, 10]. Tactics and strategies are reported by Carmel and Agarwal [15], who describe tactics for alleviating the distance between sites, by Lee and Delone [16], who performed an extensive interview study among managers on coping strategies in GSD, or by Krishna et al. [17], who focused on strategies for handling cultural differences. A few studies also specifically address work allocation in global software development and the underlying criteria [11, 12].

However, while much work has been done on analyzing the current practices of distributed development, less has been done to transfer this knowledge into approaches or methods that address the resulting challenges from a project planning or management perspective [18]. In particular, the allocation of work to sites in distributed projects has been addressed by relatively few authors, even though the problems resulting from inadequate distributions have been described well by the empirical studies cited above. Slightly more work has been published in the area of assessing future GSD projects with respect to costs and risks. In the following, we give a brief overview of the current state of research in work allocation decision support and planning and in the assessment of GSD projects.

### A. Work Allocation Decision Support

Prikladnicki et al. [19] suggest a reference model for global software development project planning. It considers two levels, strategic and tactical / operational planning of GSD projects. The key process on the strategic level is project allocation, which consists of project analysis, project distribution decision, and center (site) selection. The authors suggest a set of criteria for site selection but do not provide any systematic support for the assignment decision. Similarly, the Siemens-led Global Studio Project [20, 21] resulted in tool-supported processes and methods for assigning work to distributed teams and communicating the assignment. However, once again, little guidance is given for the specific allocation decision.

Mockus and Weiss [22] tried to directly support the decision of assigning work to different sites using a formal algorithm. They used the data from earlier modifications on the code basis of a software development project in order to optimize communication between sites by minimizing the number of modifications spanning code parts assigned to multiple sites. As this approach has a clearly defined optimization goal, it is able to identify an optimal work assignment algorithmically. However, other and more complex criteria for work allocation are not considered here.

Setamanit et al. [23, 24] developed a simulation model that can be used for evaluating different task allocation alternatives. The model contains several site-specific sub-models that reflect the special resources and capabilities at all sites. They are enhanced by an interaction effect model that uses several influencing factors to describe efficiency of communication between sites and its impact on productivity. As a result, the model can simulate the effects of work allocation strategies on project effort and time. However, the model lacks processes for using the simulation model for organization-and project-specific assignment decisions.

Sooraj and Mohapatra [25] suggest a model that uses different types of indices to evaluate different task allocation alternatives. They assume the existence of a so-called "Coordination Index" describing the overhead needed for executing a task in a distributed environment. The index depends on factors such as work coupling, time zone differences, and communication effectiveness between sites and can thus be calculated and compared for every two involved sites. This makes it possible to base an allocation decision on a quantitative evaluation of alternatives. However, it is again not specified how the model and the underlying influencing factors and quantifications can be adapted to individual environments.

### B. GSD Project Planning and Assessment

For project planning and assessment of future projects, approaches have been suggested in GSD research that regard projects from a risk assessment or cost estimation perspective.

In risk assessment, there are several approaches that name GSD-specific project risks and classify them into several categories [26], [27]. They aim at providing a generic set of experience-based threats that might have to be addressed in future projects. It is, however, left to the individual project manager to instantiate the risks for a specific GSD project. Smite [28] presents a risk identification approach that is more suited for identifying specific risks for an individual project situation as it links certain threats to project risks. A characterization of the project with respect to the threats therefore results in a project-specific evaluation of risks.

In general, most risk assessment approaches do not consider the selection of sites and the allocation of work to sites as having an influence on project-specific risks. Prikladnicki et al. [29] integrate risk assessment into the site selection process and point out the interrelationship between work allocation and project risks. Yet, they only present a generic process framework without giving specific guidance on risk assessment. Overall, current research approaches do not enough support systematic risk assessment in GSD.

In cost estimation for GSD, several extensions of COCOMO [30] have been suggested. Some approaches aim at identifying new effort multipliers that reflect the additional and complex effort overhead caused by distributed collaboration [31, 32]. However, this does not consider the impact of distributing work to several sites (with different abilities) on productivity. Madachy [33] addresses this by suggesting an individual assessment of effort multipliers for every involved site, but he does not regard the impact of work distribution on communication overhead. Therefore, these cost estimation approaches are not suited for supporting work allocation decisions by evaluating the costs of different assignment alternatives.

In general, the related work shows that approaches for decision support in work allocation are either very generic or very much narrowed down on selected criteria. The published approaches for project planning and assessment, on the other hand, do not systematically consider the impact of work allocation on project risks and costs and therefore cannot be used either for supporting work allocation. In addition, most approaches in the literature lack ways to adapt them to individual environments, experiences, or projects.

## III. THE INTEGRATED ASSIGNMENT MODEL

In the following section, we give an overview of the proposed risk-driven work allocation model.

### A. Problem and Goal

Based on our analysis of the state of the practice with respect to criteria for work allocation in GSD [12], we define the *work allocation* problem in GSD as the problem of finding an *assignment* between elements of two sets (see Figure 1):

− A set of tasks that together form the software development project. Each task can, in principle, be assigned to an individual location. Depending on the project, a task could be defined as a needed role, a process step, or the responsibility for a certain part of the product [22]. In our studies, we mostly found a division of work using a mixture of process steps and product components. For example, a project might be split up into functional design, development of several components, system test, and integration test.

− A set of sites that together form the available resources. Every task can be assigned to one site where adequate resources are available. In most situations, resources for a given task are available at several sites at the same time.

If multiple sites are available for each task, it has to be decided to which site the task should be assigned to (therefore, the number of possibilities will grow exponentially with the number of sites available). This assignment decision has to be made with respect to project-specific goals and criteria as well as the characteristics of the organizational environment. Moreover, the decision is also impacted by certain characteristics of the project (i.e., the set of tasks) and of the resources (i.e., the selected sites). Assigning very complex tasks to sites with little experience could, for example, reduce productivity and therefore might impact the project costs negatively. Assigning closely coupled tasks to sites with high time zone differences might lead to late-night shifts for the involved personnel and thus could decrease motivation (which could also be considered a project goal).

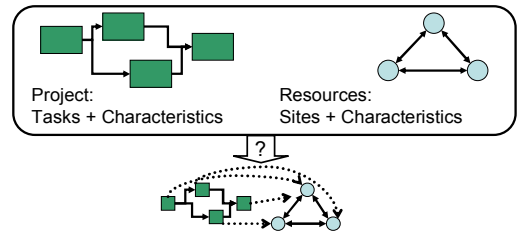

Figure 1. GSD Work Allocation

Therefore, work allocation should systematically consider an (organization-) specific set of influencing factors (in the examples above: task complexity, experience, task coupling, time zone differences), and there should also exist an understanding of how these factors might impact project goals. However, in practice, we found the following problems:

− Work allocation in GSD is neither done systematically nor transparently and relies only on the individual experiences of the project managers involved.

− Relevant criteria and influencing factors are not regarded in work allocation.

− The consequences of work allocation decisions on project goals and risks are often not considered.

In order to address these problems with a more systematic work allocation process, it has to be evaluated how different assignments impact the project goals. In addition, assignment alternatives should be identified with respect to all relevant criteria and influencing factors. Finally, the characterization of the project and the resources used as a basis for the decision should be documented together with the assignment decision in order to make it transparent and enable an organization to reuse experiences gained from previous projects for improving future decisions.

Consequently, we formulated the following goal for our work: *Develop a model for systematic work allocation in GSD projects that is able to (1) document the decision process, (2) suggest work assignments, and (3) evaluate the consequences of different alternatives. The method should be adaptable to individual contexts and based on organization-specific experiences.*

### B. Model Overview

The proposed integrated model supports two main steps of work allocation: First, an assignment is chosen from a set of alternatives based on specific characteristics of the project and the resources and their impact on individual project goals.

Then, the assignment decision is analyzed with respect to possible project risks stemming from the work distribution.

Based on a division of the project and resources into tasks and sites and a subsequent characterization of the tasks and sites, the model is able to provide a weighted list of assignment suggestions and to evaluate every alternative with respect to the expected project risks. Figure 2 gives an overview of the model input and output.

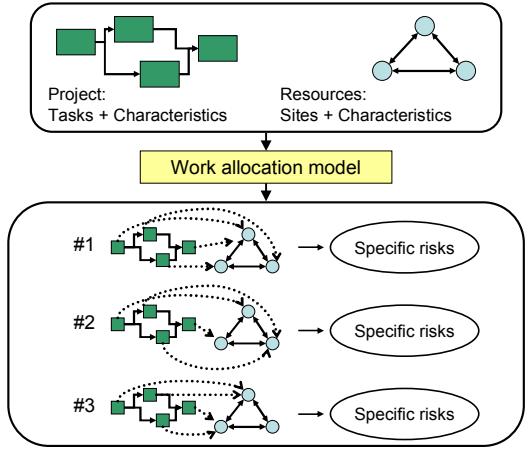

Figure 2. Model input and output.

The model consists of two main sub-models sharing a common causal model, as seen in Figure 3: The stochastic assignment model is responsible for algorithmically deriving the weighted list of assignment suggestions, while the risk identification model can predict project risks for any given assignment. Both do this using a set of influencing factors and their impact on project goals that is stored in the common causal model. The causal model is therefore able to store organization-specific experiences.

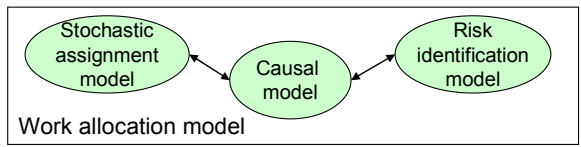

Figure 3. Basic model structure

Any new project that is supported by the model, as well as the available resources, must be characterized according to the influencing factors stored in the causal model. Afterwards, the stochastic assignment and risk identification model can be automatically executed using the stored data.

The proposed model addresses the problems and goal stated in Section 3.1 as follows:

– The causal model stores an experience-based set of influencing factors. All projects and resources are characterized according to these factors. The work allocation decision is thus made transparent and does not only rely on individual expertise.

– The assignment suggestion model provides a weighted list of alternatives based on multiple criteria and the experiences stored in the causal model.

– The risk identification model evaluates the consequences of every work assignment alternative.

In the following, the main components of the integrated model will be explained in more detail.

*1) Causal Model.* The causal model stores the organization-specific, relevant influencing factors and their impact on project goals. This is done by describing two main sets of elements: influencing factors (e.g., "language differences") and project goals (e.g., "project costs"). These items are stored as nodes in a casual network and connected via causal relationships and other nodes (e.g., "language differences" → "communication problems" → "productivity" → "project costs"). Causal relationships can be of a positive or negative nature and have different weights according to the relevance of their impact (e.g., "language differences" have a strong positive impact on "communication problems", which themselves have a medium negative impact on "productivity").

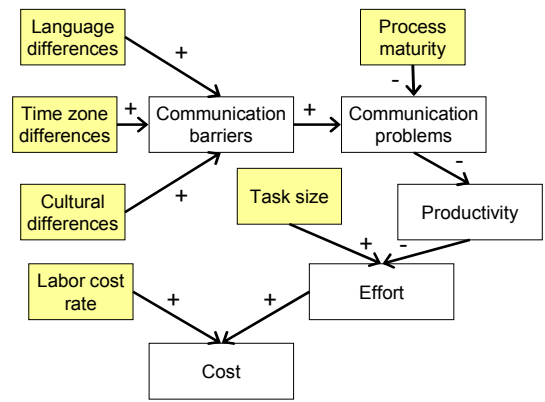

Figure 4. Simple causal model (excerpt)

Altogether, the nodes and causal relationships of the causal model capture experiences on how work distribution impacts the success of distributed software development projects in a formal way. Organization-specific experiences can be stored in the model by adding influencing factors or causal relationships and setting their weights according to the lessons learned in previous GSD projects. In a separate publication, we describe how a basic causal model was built based on a literature review and interviews with experienced practitioners [34]. An excerpt of another simple causal model in an industrial context is shown in Figure 4. The excerpt also shows how the examples above were integrated into the model.

*2) Stochastic Assignment Model.* The stochastic assignment model formalizes the work allocation decision such that it can be addressed algorithmically and at the same time captures the inherent uncertainty by applying stochastic simulation. As a result, it is able to identify a list of assignment suggestions based on the causal model and a characterization of a given project and available resources.

Internally, this is done using an algorithm from distributed systems that can identify an optimal task assignment if it is provided with defined cost functions. Therefore, the sub-model transfers the project characterization and the causal model into numeric cost functions of the underlying algorithm. However, the inherent uncertainty in modeling human-based projects cannot be adequately captured by simple numeric cost functions. Instead, we chose Bayesian networks as a probabilistic technique for modeling the GSD project. This method has been used several times in the context of planning and managing software development projects [35]. In the stochastic model, the Bayesian networks aggregate the impact of work distribution on project goals into a probabilistic distribution for the underlying cost functions [13].

Altogether, the stochastic assignment sub-model performs the following steps:

1) The causal model is transformed into Bayesian networks.

2) Using the characterization of the project and the resources, the impact of every possible assignment on project goals is inferred in the Bayesian networks. The impact is aggregated based on weighted project goals and stored as probabilistic distributions for the cost functions needed in the task assignment algorithm.

3) In a large number (e.g., 1000) of runs, the cost functions are instantiated based on the probabilistic distributions and the assignment algorithm from distributed systems is executed.

4) For every run, the returned optimal assignment is stored. All returned assignments are then ordered by their number of appearance (i.e., the number of runs in which the assignments were returned). This is finally presented as a weighted list of assignment suggestions.

Steps 2-4 are described in more detail in a different publication [13] that proposes a standalone version of the stochastic assignment model. However, as the stochastic assignment model is now integrated with the risk identification model, the Bayesian networks are built from the common causal model. In order to do so, the probabilistic tables in the Bayesian networks have to be quantified.

*3) Risk Identification Model.* The risk identification model is able to predict GSD-related risks for a given project and work allocation. Therefore, it can be used for analyzing and comparing the different assignment alternatives suggested by the application of the stochastic assignment model. The model does this by transferring lessons learned into a set of semi-formal logical rules.

Every rule of the risk identification model describes how certain factors can cause GSD-specific problems [14]. Therefore, a rule is formulated as "cause → problem", with "cause" being a logical combination of influencing factors and "problem" being a GSD-specific problem that negatively affects project goals. One example of a rule could be:

*"(Cultural differences) & ¬(common working history) → communication problems"*

which describes the experience that cultural differences between two sites lead to communication problems (which negatively affect productivity) if the two sites have no history of working together (and thus have not had the opportunity to get used to their mutual differences).

As in the stochastic assignment model, the influencing factors stem from the causal model and thus are used for characterizing a project and the available resources. With such a project-specific evaluation of the influencing factors, the relevance of every rule can be evaluated for each possible assignment of tasks to sites. Therefore, any given assignment in a specific project can be analyzed with respect to the risks described in the risk model and the results can be used for comparing different assignment alternatives or for enacting countermeasures in a given assignment.

Due to its low complexity, the model can be easily applied by practitioners for documenting lessons learned and experiences from previous projects in a semi-formal and readable way.

*C. Organization-specific Instantiation*

The set of relevant influencing factors and their impact on project goals can differ from one software development organization to the next. Therefore, we recommend to instantiate the common causal model and the sub-models individually for every organization.

In order to do so, the lessons learned stored in the risk identification model can be used for deriving an initial causal model: The full set of rules documented in the risk identification model results in a set of influencing factors that provide input for the causal model. In addition, the logical combination of influencing factors and problems in the rules describes an initial set of causal relationships between influencing factors, problems, and project goals. Figure 5 demonstrates how the exemplary rule described above can automatically be transferred into an initial causal model for a Bayesian network.

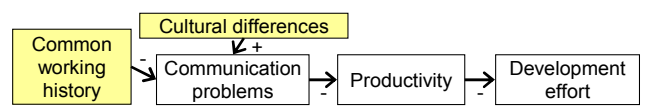

Figure 5. Causal model created from a risk model

In order to derive a complete causal model and the Bayesian networks used in the stochastic assignment sub-model, further steps have to be performed:

1) Causal mode enhancement: The initial causal model has to be enhanced by adding additional influencing factors, goals, and causal relationships. This must be done manually by experienced experts of the organization. However, as the causal model is relatively simple, it can be done without requiring much effort for familiarization. The causal model enhancement is finished if all relevant project goals and influencing factors of the organization are contained in the model.

2) Creation of the Bayesian networks: The enhanced causal model has to be transformed into Bayesian networks. We developed an algorithm that performs this transformation automatically.

3) Quantification of the Bayesian networks: The Bayesian networks have to be quantified by setting values to the probabilistic table of every node. In accordance with established tools for creating Bayesian networks [36], we defined a set of standard functions (e.g., minimum, maximum, and weighted mean) that can be used for defining the causal relationships between nodes and that are automatically transformed into probabilistic tables.

## IV. MODEL DEVELOPMENT AND APPLICATION PROCESS

In the following, we will sketch a process for development and application of the assignment model. Figure 6 gives an overview of both processes.

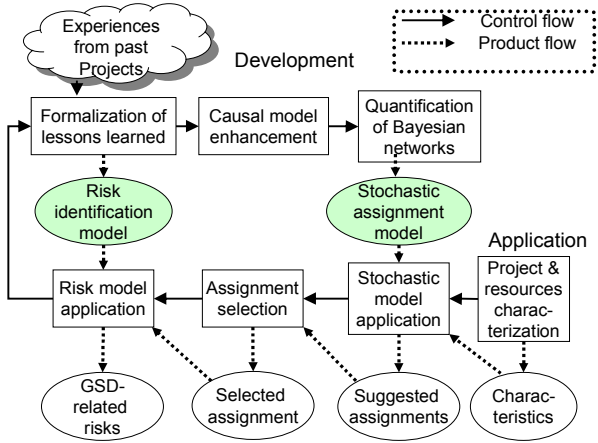

Figure 6. Model development and application

### A. Model Development

In order to develop an organization-specific allocation model that is based on custom experiences, three tool-supported steps have to be performed within every organization: First, lessons learned from previous projects concerning risks and problems in distributed software development are collected. These lessons learned can be collected in project touchdown meetings, group discussions, or interviews with individual experts. Other sources of information can be published experiences and empirical studies. These lessons learned are documented in a semi-formal way as logical rules. Therefore, for each lesson learned, influencing factors and problems have to be defined.

Afterwards, the initial causal model created from the documented lessons learned is enhanced and the algorithmically created Bayesian networks are quantified as described above.

### B. Model Application

The application of the model during project planning is done during the following process steps, which are again tool-supported:

1) Project and resources characterization: Tasks and available sites of the project are defined, as is the resource availability at the different sites. The project is characterized according to the influencing factors contained in the stochastic assignment model. Factors that address the complete project (e.g., time pressure) are evaluated only once; factors that address characteristics of tasks (e.g., complexity) or sites (e.g., experience level) or relationships between tasks (e.g., coupling) or sites (e.g., cultural differences) are evaluated individually for each task, site, or relationship between tasks or sites. In accordance with other project planning models such as COCOMOII, this evaluation is done on an ordinal scale (e.g., "very low" – "very high").

2) Stochastic model application: The stochastic model is applied by instantiating the Bayesian networks and executing the underlying assignment algorithm in multiple runs. This is done automatically and results in a weighted list of assignment suggestions.

3) Assignment selection: The decision maker analyzes the list of suggested assignments and selects one assignment. In order to support this step, the risk identification model can automatically identify specific risks for every assignment alternative. This is done by evaluating every rule of the risk model with respect to the influencing factor characterization done in the first step.

4) Risk model application: Finally, the risk identification model is applied to the selected assignment and the project-specific relevant risks are documented individually for each site in order to enact countermeasures during project management.

After project execution, the experiences gained in the project can be used for revising and adapting the model by adding new rules or altering rules in the risk identification model, changing the influencing factors or causal relationships in the causal model, or changing the probabilistic tables in the Bayesian networks of the stochastic assignment model.

## V. EVALUATION

In the following, we describe an evaluation of the approach in an industrial context. The evaluation considers the assignment suggestions of the integrated model. The risk identification capabilities were evaluated in a previously performed study [14], which revealed that more than 80% of the predicted risks actually occurred in the analyzed projects and that 40% of these had not been considered at project start.

### A. Study Goal and Research Questions

For the evaluation of the assignment suggestions, we stated the following hypothesis:

*The stochastic assignment model, when applied to past projects, is able to make assignment suggestions that make sense for experienced managers from these projects.*

Based on this hypothesis, we developed a prototype instantiation of the stochastic assignment model and formulated the following research questions:

**RQ1**: Do the suggestions made by the model match the actual work distribution in an industrial project?

**RQ2**: Do the suggestions made by the model match the retrospective view of experienced project managers?

**RQ3**: Do the experienced managers consider the suggestions made by the model to be reasonable and helpful?

*B. Design, Context, and Execution*

The evaluation was done via a series of semi-structured interviews intended to analyze distributed software development projects in retrospective. We decided to use historic projects as unit of analysis because they offer the possibility to reconstruct both the work allocation decision and its impact on project goals. Interviews were held with experienced managers from the analyzed projects.

The study was conducted at Indra Software Labs (ISL) in Madrid, Spain. Indra is the largest IT company in Spain and a leading IT multinational in Europe. ISL is the network of Software Labs of Indra that develops customized software solutions for Indra's markets. It has 20 development sites, half of which are located in Spain and the others in Latin America, Slovakia, and the Philippines. Most of the software development projects at ISL are distributed either within Spain or globally. Therefore, within ISL there exists a lot of experience regarding work in GSD projects and related risks and problems.

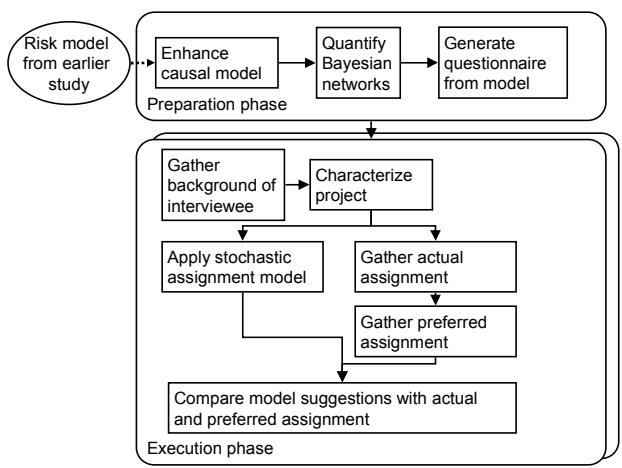

Figure 7. Evaluation process.

We divided the evaluation into a preparation and an execution phase (see Figure 7). In the evaluation phase, a work allocation model was built based on the processes described in Sections 4.1 and 3.3: The risk model developed (and successfully evaluated) in an earlier study at ISL [14] was used as input. From this model, a causal model was derived and enhanced based on experiences gathered in earlier literature and interview studies [12, 34]. An excerpt of the resulting causal model can be seen in Figure 4. The automatically derived Bayesian networks of the stochastic assignment models were again quantified based on the data gathered in the previous studies. All models were inserted into a Java tool that implements the model. Finally, a questionnaire for project characterization was generated using the influencing factors of the causal model. This questionnaire was then sent to the interviewees in advance, asking them to select one historic project they were involved in (ideally with a complex work allocation decision) and characterize it.

In the execution phase, each interview was conducted separately according to the following process: First, the interviewee was questioned about his background (position and experience in GSD). Then, he was asked to characterize the tasks and sites of the chosen project with respect to the influencing factors of the used model. In parallel to his answers, the characterization was inserted into the model implementation. This made it possible to apply the stochastic model to the project during the interview. During the execution of the stochastic model (but before the results were presented), the interviewee was asked to name the actual allocation of tasks to the different sites as well as the preferred assignment in his retrospective view. Afterwards, the work allocation suggested by the model was compared with the actual and preferred assignment and the interviewed manager was asked to comment on that.

Overall, four projects were analyzed in July and August 2010. Each project was analyzed in an interview lasting one hour. One interview was done over the telephone while the others were conducted in video-conferences. In addition to the project characterization stored in the Java tool, further detailed notes were taken during the interview.

*C. Results*

All interviewed persons could report from several years of experience as a project manager or a manager of a local ISL center. They were all experienced with GSD, having participated in 2-10 distributed projects. In the following, the results of each project evaluation are described. All projects developed information systems for external customers.

**Project A**. This project was done at centers in Madrid and La Coruna. The work consisted of function design and the individual development of three modules: Two modules (A and B) managed specific sub-components, while the third module (C) was responsible for central data management. The Madrid site had a higher cost rate than La Coruna but could interact more easily with the customer, who was also located in Madrid. In the actual project, functional design and module C were assigned to the Madrid site, whereas modules A and B were done in La Coruna. However, the project manager reported that she would rather have done everything in La Coruna, even though she admitted that the functional design would probably have to be done in Madrid due to the customer's location. The stochastic assignment model made a similar suggestion: The highest-ranked alternative was assigning everything to La Coruna, followed by an assignment of functional design to Madrid and development to La Coruna.

**Project B**. The project was managed at an ISL site in Lleida (Spain). This site had previously started cooperating with an Argentinean site of ISL which it could use for development. The customer was located in Barcelona where an additional ISL center is located. The project was subject to a high time pressure (the product had to be in use within 9 months) and unstable requirements. It covered all phases of the lifecycle: Functional design, technological design, development, and integration testing had to be assigned to the three sites. In the actual project, technological design was done together with project management in Lleida; development was assigned to Argentina; functional design and integration testing were done in Barcelona near the customer. The stochastic model did not suggest this assignment. Instead, its first two suggestions were to do everything in Lleida with integration testing either in Barcelona (#1) or in Lleida as well (#2). Only the fourth suggestion in the ordered list contained an assignment of development to Argentina. Confronted with the suggestions of the model, the project manager stated that it was their first idea to assign everything to Lleida (as in the second suggestion of the model), but finally it was decided to use Barcelona and Argentina as well. He also reported that including the Argentinean site indeed complicated project management. However, in the project this could be alleviated by sending a person from Spain to Argentina to help manage the interface between the sites.

**Project C**. In the project coordinated from the Madrid center of ISL, the customer was in Portugal where ISL does not have any sites. Other sites involved in the project were Valencia, Ciudad Real, and Sevilla. The project was split up into functional design, an enrichment of the functional design with special market knowledge, and the development of three components of equal size. The actual project assigned functional design to Madrid, enrichment of the design to Valencia, and the three components to Valencia, Ciudad Real, and Sevilla. However, the model did not have this assignment in its suggestions. Instead, it suggested first to assign everything to Valencia, followed by an assignment of both functional design tasks to Sevilla and development to Ciudad Real. This was similar to the opinion of the project manager who stated that in his view, the project should have been done at one location. In particular, he pointed out that the two design tasks in Madrid and Valencia had many collaboration problems. The model reflected this, as it assigned these two tasks to one single site in all suggestions.

**Project D**. The requirements specification of this project had to be assigned to Valencia. As development was done in COBOL, it could only be assigned to sites with teams available that could program in COBOL. These were in Salamanca, Badajoz, and a site in Argentina. In the actual project, the development was assigned to Badajoz and Argentina. The model, however, suggested an assignment only to Badajoz or Salamanca first, followed by suggestions to split the work between either Badajoz or Salamanca and Argentina. However, if the parameters of the underlying causal model were adapted to the views of the project manager (he saw additional gains in work distribution such as knowledge transfer that were not covered in the model), suggestions to split up the work between Spain and Argentina were ranked higher.

Based on the project evaluation, we can answer the research questions as follows:

**RQ1**: No – In none of the projects, the actual work distribution matched the distribution suggested by the model with the highest rank. Often, this was caused by not enough available personnel at the sites suggested by the model or by political decisions.

**RQ2:** Mostly – In projects A and C, the model suggestions matched the opinions of the involved project manager. In project B, the model suggestion did not comply with the opinion of the manager but the suggestion had been considered first in the assignment process and had been favored by other project managers. In project D, the manager suggested an assignment different from the one suggested by the model, but these differences could be reduced by adjusting the quantifications in the underlying Bayesian network.

**RQ3:** Yes - In projects A and C, the fact that the model complied with the retrospective opinions of the managers shows that the managers saw the suggestion as reasonable. In addition, the difference between suggested and actual assignment demonstrates that the use of the model would have been helpful, which was also confirmed by the managers. In project B, the suggestion made by the model actually was discussed during the decision process and thus was seen as reasonable. In project D, the use of the model helped to make the underlying assumptions regarding the risks and benefits of work distribution explicit.

TABLE I. RESEARCH QUESTIONS 1 AND 2 FOR EACH PROJECT

| Project | Suggested = actual distribution? | Suggested distribution = manager's view? | Suggested distribution reasonable? |
|---|---|---|---|
| A | No | Yes | Yes |
| B | No | Partly[1] | Yes[1] |
| C | No | Yes | Yes |
| D | No | No[2] | Partly[2] |
| 1: Suggested distribution was first idea in decision process [2]: Different assumptions on the benefits of distribution; change of causal model resulted in compliance ||||

Table 1 gives an overview of how the research questions can be answered for the different projects.

*D. Threats to Validity*

Internal validity might be threatened by a lack of common understanding with respect to the influencing factors: Both the risk model and the stochastic assignment model characterized the projects with respect to a set of influencing factors. There might be different interpretations among the project managers about the exact meaning of each factor. We tried to reduce this threat by explaining every factor in detail if necessary. Another threat is given by the fact that we used the opinion of the experienced project managers for judging the quality of the model, even though the preferred distribution might not have been optimal for the project. We mitigated this risk by specifically asking for historic projects the managers were involved in. Therefore, they had very good insight into the analyzed projects and also knew about the outcomes of the work allocation actually used. Additionally, validity might be threatened by the fact that the evaluation was conducted by the same person who developed the model. Another threat could have been that the same experiences were used for model construction and validation. This risk was mitigated by performing the evaluation with different managers than the model development.

External validity might be threatened by the fact that the evaluation was done at only one company with specific characteristics. For example, all analyzed projects reported on distributed development within Spain or between Spain and Latin America, which is due to the fact that most development centers of ISL are located in these two regions. This implies that the model could not be evaluated in projects with large language or cultural differences. Therefore, the study should be repeated at different organizations.

## VI. CONCLUSION AND FUTURE WORK

In this article, we proposed a risk-driven model for supporting work allocation decisions in global software development projects that consists of two main components: A stochastic assignment model (1), which combines a task assignment algorithm from distributed systems with stochastic simulation using Bayesian networks in order to provide a project-specific list of assignment suggestions. A risk identification model (2), which can evaluate the risks of a given assignment in a specific project based on previous lessons learned. Both sub-models communicate over a common causal model and collaborate twofold: On the one hand, the suggestions made by the stochastic assignment model can be evaluated using the risk identification model. On the other hand, the formalized lessons learned from the risk identification model can be used as input for the development of an organization-specific and experience-based stochastic assignment model.

The evaluation of the model demonstrated that it is able to address the first two of the problems stated in Section 3.1: In all analyzed projects, it was possible to model the breakdown of the project into distinct tasks and to characterize both tasks and sites according to the relevant influencing factors stored in the causal model. This characterization, together with the model suggestions and the selected assignment, can be stored as an xml file and thus makes the work allocation decision transparent and documented. Using the process described in Section 4.1, it was also possible to create an organization-specific assignment model that is able to make reasonable suggestions and therefore can systematically apply previous experiences to future allocation decisions. Furthermore, the evaluation showed that the model is able to make assignment suggestions that consider multiple criteria and that often comply with the retrospective view of experienced project managers. Even if the suggestions did not always comply, they could be used as a basis for a discussion in all analyzed projects. In projects A and C, the model did comply with the retrospective view of the manager, which was different from the actual project distribution. We argue that the actual use of the models during project planning could have changed the allocation decision and led to higher productivity.

As the ability to address the third problem stated in Section 3.1 has already been evaluated separately using the risk identification model [14], we see the proposed risk-driven allocation model as a way to improve GSD project planning towards more systematic allocation decisions that increase the success rate of globally distributed software development projects.

However, the approach also has some limitations that might prevent its applicability in some industrial contexts: The approach requires some upfront effort in order to develop organization-specific models and characterize upcoming projects with respect to the model parameters (in our case studies, this upfront effort included conducting and analyzing 10 to 15 interviews as well as approximately two days for model building). Even though we believe that this effort is very much paid back by the improvements in project planning and work distribution, which can result in higher productivity, this might be a barrier to the application of the model. In addition, not all projects are defined well enough in order to be broken down into distinct tasks that can be characterized in the model (e.g., agile projects). Therefore, open research questions remain in terms of applicability in different environments which we plan to address in several future evaluation studies.

## ACKNOWLEDGEMENTS

The authors would like to thank all participants in the interview and evaluation studies. At ISL, these persons were: Josep Fernández González, Miguel Ángel Ballesteros Velasco, Miguel Rovisco Pais, Maria Jesus Pérez González, Diego de Alcalá Cachero Rodréguez, Antonio Fernández Zaragoza, Ramon Lema García and David Alonso Fernández. The authors also thank Sonnhild Namingha for proofreading the paper.

## REFERENCES

[1] Damian, D. and Moitra, D. 2006. Global software development: How far have we come? *IEEE Software*, 23, 5 (Sept. 2006), 17-19. DOI= http://dx.doi.org/10.1109/MS.2006.126

[2] Herbsleb, J. D. and Mockus, A. 2003. An Empirical Study of Speed and Communication in Globally-Distributed Software Development. *IEEE Trans. Software Engineering*. 29, 6, (Jun. 2003), 481-494. DOI= http://dx.doi.org/10.1109/TSE.2003.1205177

[3] Herbsleb, J. D., Paulish, D. J., and Bass, M. 2005. Global software development at Siemens: Experience from nine projects. *Proceedings of the International Conference on Software Engineering*. ICSE '05. 524-533. DOI=http://dx.doi.org/10.1145/1062455.1062550

[4] Smite, D. and Moe, N. B. 2007. Understanding a Lack of Trust in Global Software Teams: A Multiple-Case Study. *Proceedings International Conference on Product Focused Software Development and Process Improvement*. PROFES '07. 20-34. DOI=http://dx.doi.org/10.1007/978-3-540-73460-4_6

[5] Bass, M. and Paulish, D. 2004. Global software development process research at Siemens. *Proceedings of the Third International Workshop on Global Software Development* at International Conference on Software Engineering, ICSE '04. 8-11.

[6] Heeks, R., Krishna, S., Nicholson, B., and Sahay, S. 2001. Synching or Sinking: Global Software Outsourcing Relationships. *IEEE Software*. 18, 2, (Mar. 2001). 54-60. DOI=http://dx.doi.org/10.1109/52.914744.

[7] Casey, V., and Richardson, I. 2006. Uncovering the Reality within Virtual Software Teams. *Proceedings of the International Workshop on Global software development for the practitioner*, 2006. 66-72. DOI= http://dx.doi.org/10.1145/1138506.1138523

[8] Casey, V. 2009. Leveraging or Exploiting Cultural Difference? *Proceedings of the International Conference on Global Software Engineering* ICGSE '09. 8-17. DOI=http://dx.doi.org/10.1109/ICGSE.2009.9

[9] Carmel E. and Abbott, P. 2007 Why 'nearshore' means that distance matters. *Commun ACM*. 50, 10 (Oct. 2007). 40-46. DOI=http://dx.doi.org/10.1145/1290958.1290959

[10] Milewski, A. E., Tremaine, M., Egan, R., Zhang, S., Kobler, F., and O'Sullivan, P. 2008. Guidelines for Effective Bridging in Global Software Engineering. *Proceedings of the International Conference on Global Software Engineering*, ICGSE '08. 23-32. DOI=http://dx.doi.org/10.1109/ICGSE.2008.16

[11] Edwards, H. K, Kim, J. H., Park, S., and Al-Ani, B. 2008. Global software development: Project decomposition and task allocation. *International Conference on Business and Information*, BAI '08. Academy of Taiwan Information Systems Research, ISSN: 1729-9322

[12] Lamersdorf, A. Münch, J., and Rombach, D. 2009. A survey on the state of the practice in distributed software development: Criteria for task allocation. *Proceedings of the International Conference on Global Software Engineering*, ICGSE '09. 41-50. DOI=http://dx.doi.org/10.1109/ICGSE.2009.12

[13] Lamersdorf, A. and Münch, J. 2010. A multi-criteria distribution model for global software development projects. *J Braz Comp Soc*. 16 (2010). 97-115. DOI=http://dx.doi.org/10.1007/s13173-010-0010-6

[14] Lamersdorf, A. Münch, J., and Rombach, D. 2010. A Rule-based Model for Customized Risk Identification in Distributed Software Development Projects. *Proceedings of the International Conference on Global Software Engineering*, ICGSE '10

[15] Carmel, E. and Agarwal, R. 2001. Tactical Approaches for Alleviating Distance in Global Software Development. *IEEE Software*, 18, 2 (Mar. 2001), 22-29. DOI=http://dx.doi.org/10.1109/52.914734

[16] Lee, G., DeLone, W., and Espinosa, J. A. 2006. Ambidextrous coping strategies in globally distributed software development projects. *Comm. ACM*, 49, 10 (Oct. 2006), 35-40. DOI=http://dx.doi.org/10.1145/1164394.1164417

[17] Krishna, S., Sahay, S., and Walsham, G. 2004. Managing cross-cultural issues in Global Software Outsourcing. *Comm .ACM*, 47, 4 (Apr. 2004), 62–66. DOI=http://dx.doi.org/10.1145/975817.975818

[18] Herbsleb, J. D. 2007. Global software engineering: The future of socio-technical coordination. *Proceeding of the Future of Software Engineering*, FOSE 2007, 188-198. DOI=http://dx.doi.org/10.1109/FOSE.2007.11

[19] Prikladnicki, R., Audy, J. L. N., and Evaristo, R. 2006. A Reference Model for Global Software Development: Findings from a Case Study. *Proceedings of the International Conference on Global Software Engineering*, ICGSE '06. 18-28. DOI=http://dx.doi.org/10.1109/ICGSE.2006.261212

[20] Mullick, N., Bass, M., Houda, Z., Paulish, D. J., Cataldo, M., Herbsleb, J. D., and Bass, L. 2006. Siemens Global Studio Project: Experiences Adopting an Integrated GSD Infrastructure. *International Conference on Global Software Engineering*, ICGSE '06, 203-212. DOI=http://dx.doi.org/10.1109/ICGSE.2006.261234

[21] Raghvinder, S., Bass, M., Mullick, N., Paulish, D. J., and Kazmeier, J. 2006. *Global Software Development Handbook*. Auerbach Publications, London, 2006

[22] Mockus, A., and Weiss, D. M. 2001. Globalization by Chunking: A Quantitative Approach. *IEEE Software*, 18, 2, (March 2001), 30-37. DOI=http://dx.doi.org/10.1109/52.914737

[23] Setamanit S., Wakeland W. W., and Raffo D. 2007 Using simulation to evaluate global software development task allocation strategies. *Software Process: Improvement and Practice*, 12, 5, 491-503. DOI=http://dx.doi.org/10.1002/spip.335

[24] Setamanit S, and Raffo D (2008) Identifying key success factors for globally distributed software development project using simulation: A case study. *Proceedings of the International Conference on Software Process*, ICSP'08. 320-332. DOI=http://dx.doi.org/10.1007/978-3-540-79588-9_28

[25] Sooraj P., and Mohapatra P. K. J. 2008. Developing an Inter-site Coordination Index for Global Software Development. *Proceedings of the International Conference on Global Software Development*, ICGSE'08. 119-128. DOI=http://dx.doi.org/10.1109/ICGSE.2008.30

[26] Ralyte, J., Lamielle, X., Arni-Bloch, N., and Leonard, M. 2008. A framework for supporting management in distributed information systems development. *Proceedings of the International Conference on Research Challenges in Information Science*, RCIS '08. 381-392. DOI= http://dx.doi.org/10.1109/RCIS.2008.4632128

[27] Ebert, C., Murthy, B. K., and Jha, N. N. 2008 Managing risks in global software engineering: principles and practices, *Proceedings of the International Conference on Global Software Engineering*, ICGSE '08, 131-140. DOI=http://dx.doi.org/10.1109/ICGSE.2008.12

[28] Smite, D. 2007. Project outcome predictions: Risk barometer based on historical data. *Proceedings of the International Conference on Global Software Engineering*, ICGSE '07, 103-112. DOI=http://dx.doi.org/10.1109/ICGSE.2007.37

[29] Prikladnicki, R., Yamaguti, M. H., and Antunes, D. C. 2004. Risk management in distributed software development: A process integration proposal, *Proceedings of the IFIP Working Conference on Virtual Enterprises*, 2004, 517-526.

[30] Boehm, B., Abts, C., Brown, A., Chulani, S., Clark, B., Horowitz, E., Madachy, R., Reifer, D., and Steece, B. 2000. *Software Cost Estimation with COCOMO II*. Prentice-Hall, New Jersey, 2000

[31] Betz, S. and Mäkiö, J. 2008. Applying the OUTSHORE approach for risk minimisation in offshore outsourcing of Software Development projects. *Proceedings of the Multikonferenz Wirtschaftsinformatik*, 2008, 1101-1112.

[32] Keil, P., Paulish, D. J., and Sangwan, R. 2006. Cost estimation for global software development. *Proceedings of the International Workshop on Economics Driven Software Engineering*, 2006, 7–10. DOI=http://dx.doi.org/10.1145/1139113.1139117

[33] Madachy, R. 2007. Distributed global development parametric cost modeling. *Proceedings of the International Conference on Software Process*, ICSP '07, 159-168. DOI=http://dx.doi.org/10.1007/978-3-540-72426-1_14

[34] Lamersdorf, A. and Münch, J. 2010 Studying the Impact of Global Software Development Characteristics on Project Goals: A Causal Model. *The Open Software Engineering Journal*, 4, (2010), 2-13. DOI=http://dx.doi.org/10.2174/1874107X01004020002

[35] Fenton, N., Marsh, W., Neil, M., Cates, P., Forey, S., and Tailor, M. 2004. Making Resource Decisions for Software Projects. *International Conference on Software Engineering*, ICSE'04, 397-406.

[36] Agena Limited. *AgenaRisk Tool*. http://www.agenarisk.com/products/ (Retrieved 07/23/2010f)